# Iris-assisted Terahertz Field-Induced Second Harmonic Generation in Air

Amit Beer[1,2], Dror Hershkovitz[1,2] and Sharly Fleischer[1,2]*

[1]Raymond and Beverly Sackler Faculty of Exact Sciences, School of Chemistry, Tel Aviv University 6997801, Israel.
[2]Tel-Aviv University center for Light-Matter-Interaction, Tel Aviv 6997801, Israel.
*Corresponding author: sharlyf@post.tau.ac.il

**Abstract:** Terahertz field-induced second harmonic generation (TFISH) is a technique for optical detection of broad-band THz fields. We show that by placing an iris at the interaction volume of the THz and optical fields, the TFISH signal increases by few ten-fold in atmospheric air. The iris-assisted TFISH amplification is characterized at varying air pressures and probe intensities and provides an elegant platform for studying nonlinear phase-matching in the gas phase.

________________________________________________________________________________________

### Introduction:

Terahertz fields with frequencies $10^{11}$ - $10^{13}$ Hz have become table-top available in the last two decades and are widely utilized in various fields of research and technology [1–4]. THz spectroscopy is typically performed in time-domain spectrometers consisting of a THz source (photoconductive antenna [5] / two-color plasmas [6] / rectification in a non-linear crystal [7–9] and recently also in surfaces patterned with metamaterials [10,11]) and a time-resolved THz detection module such as photoconductive antenna receiver, electro-optic sampling [12,13] or TFISH [14,15]. The latter, termed and utilized by Cook et.al. for studying THz-induced dynamics of liquid water [14] was explored and developed extensively by Zhang and company to provide optical characterization of broad-band THz fields in ambient air and various gasses [16–19]. Later on, the same group developed the air-biased-coherent-detection technique (ABCD) [20] for heterodyne detection of broad-band THz fields. The development and optimization of TFISH as a viable method for THz detection yielded various observations such as the effect of Gouy phase of both the THz and the optical probe [21], effects of different gas molecules [22] and gas densities [23], TFISH dependence on the optical probe intensity [17] and many others [24].

The non-linear optical effect that governs TFISH is the mixing of three input fields – a THz field ($E_{THz}$) and two optical fields ($E_{\omega 1}, E_{\omega 2}$) via the 3rd order susceptibility $\chi^{(3)}$ to yield a signal field at the frequency of $\omega_{signal} = \omega_1 + \omega_2 \pm \omega_{THz}$. Typically performed with one optical beam (hence $\omega_1 = \omega_2 \equiv \omega$) and since $\omega_{THz} \ll \omega$, the TFISH signal is observed at $\omega_{TFISH} \cong 2\omega$, similar to EFISH (Electric-field induced second harmonic generation [25]) only with $\omega_{THz}$ replacing for the dc electric field. The signal field amplitude is given by: $E_{2\omega} \propto \chi^{(3)} E_{THz} E_{\omega}^2$ (1) and signal intensity $I_{2\omega} \propto {\chi^{(3)}}^2 I_{THz} I_{\omega}^2$ is linear and quadratic with the THz and optical intensities respectively.

### Experimental:

The setup used in this work is shown in Figure 1 and includes a THz field (pump, $E_{THz}$) generated via tilted-pulse-front optical rectification in a LiNbO$_3$ (Lithium-Niobate) prism [26] and a short optical probe pulse (800nm, $E_{\omega}$) routed through a computer-controlled delay stage. Both fields are collinearly focused into a static-pressure gas cell and the signal at 400nm is measured by photomultiplier and lock-in detection (Stanford Research Systems SR830) at the modulation frequency of the THz beam (500Hz).

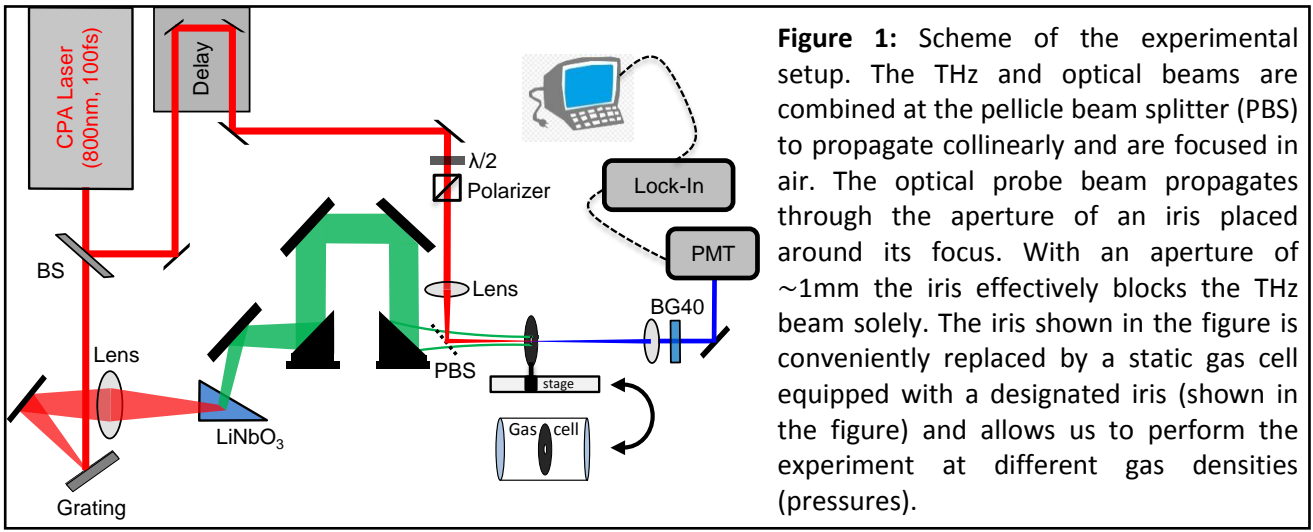

**Figure 1:** Scheme of the experimental setup. The THz and optical beams are combined at the pellicle beam splitter (PBS) to propagate collinearly and are focused in air. The optical probe beam propagates through the aperture of an iris placed around its focus. With an aperture of ~1mm the iris effectively blocks the THz beam solely. The iris shown in the figure is conveniently replaced by a static gas cell equipped with a designated iris (shown in the figure) and allows us to perform the experiment at different gas densities (pressures).

With our 100fs probe pulse energy kept below $60\mu J/pulse$ and a focal length $f = 250mm$ (with diameter of $D = 10mm$ at the focusing lens), the optical probe intensity $I_\omega < 1.1 \times 10^{14} \, W/cm^2$ is well-within the incoherent detection region noted in [17], giving rise to a unipolar, homodyne-detected signal.

Figure 2a shows our experimental TFISH intensity in ambient lab atmosphere as a function of the probe intensity. The experimental results show that by simply placing an iris at the focus of the optical probe, the TFISH intensity increases dramatically (red circles). In fact, in our setup we can hardly detect the TFISH signal obtained with probe intensities $I_\omega < 10^{13} W/cm^2$ without the iris in position (blue x's), while obtaining few ten-fold enhancement in signal intensity upon positioning the iris.

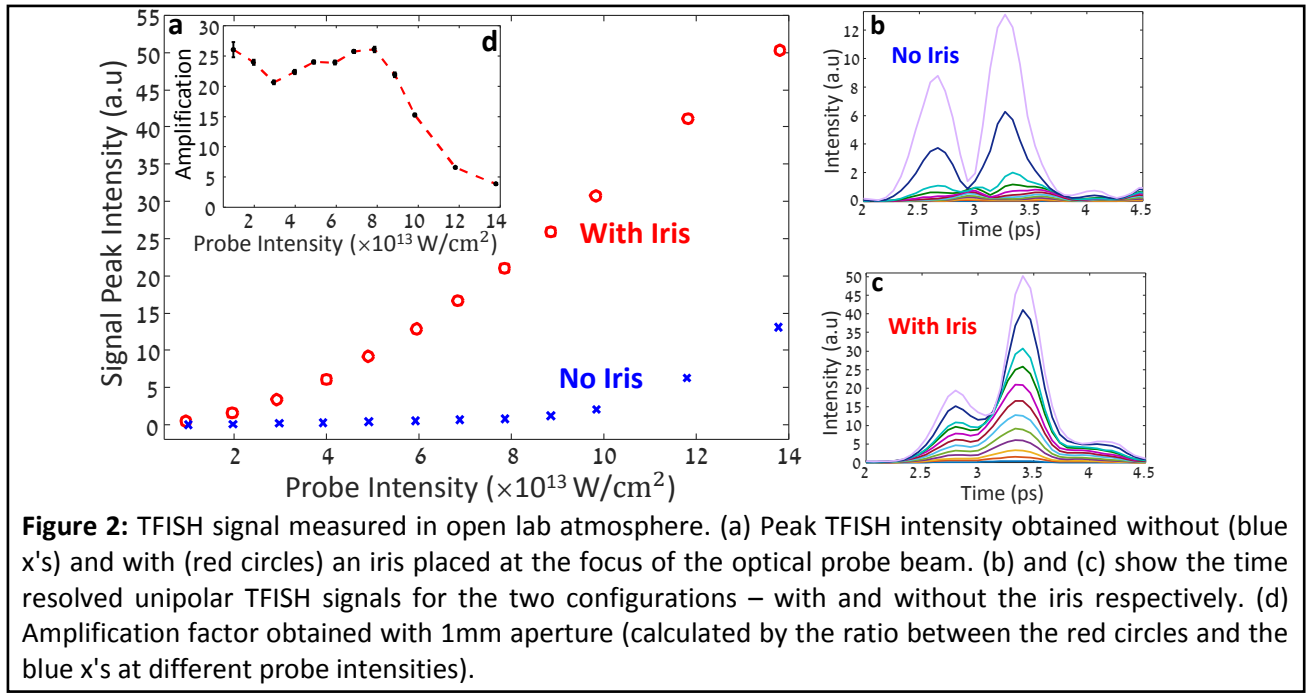

**Figure 2:** TFISH signal measured in open lab atmosphere. (a) Peak TFISH intensity obtained without (blue x's) and with (red circles) an iris placed at the focus of the optical probe beam. (b) and (c) show the time resolved unipolar TFISH signals for the two configurations – with and without the iris respectively. (d) Amplification factor obtained with 1mm aperture (calculated by the ratio between the red circles and the blue x's at different probe intensities).

Figure 2d depicts the amplification factor (namely the ratio of the red/blue signal intensities) showing a fairly fixed ~24 fold amplifications for probe intensity $I_\omega < 8 \times 10^{13} W/cm^2$. At higher probe intensities the TFISH signal increases even without the iris and the amplification factor reduces correspondingly. The time-resolved TFISH signals with and without the iris are shown in Figs. 2b and 2c respectively, with a unipolar shape as previously reported in [17] for the incoherent detection region.

In what follows we explore the dramatic increase in TFISH signal upon placement of the iris at the focus of the optical pump. Since the waist of the probe beam at the focus ($w_{probe} < 50 \mu m$) is much smaller than the aperture ($D_{aperture} > 250 \mu m$), the probe pulse is fully transmitted through the aperture (also verified by comparing the power of the probe beam at the upstream and downstream direction of the iris with no apparent difference). Next, we measured the THz power transmitted through the closed iris using a pyroelectric-detector (Gentec THZ9B-BL-DZ) and found <5% transmission owing to the large THz beam waist (~4mm diameter at the focus). Thus, in what follows we consider the aperture as fully transmitting for the optical probe and as fully blocking the THz beam for simplicity.

### Iris properties

Puzzled by the experimental results of Fig.2, we set to characterize the dependence of the iris-assisted amplification on the diameter of the aperture. First we considered the possibility of THz reflection from the metal surface of the iris that could result in local enhancement of the THz field at the upstream direction from the iris. This possibility was dismissed by replacing the flat metal aperture by a conical metal skimmer (1.3mm aperture diameter) that reflects and scatters the THz field away from the interaction volume and resulted in similar amplification as for the flat metal iris.

Next, we characterized the iris-amplification with varying aperture diameters and different probe intensities. Five metal irises were fabricated at the TAU chemistry machine shop with aperture diameters of 0.3mm, 0.5mm, 0.7mm, 1mm and 1.3mm all with 0.2mm width. The experimental amplification factors are shown in Figure 3 for three different probe intensities $2 \times 10^{13}, 4 \times 10^{13}, 6 \times 10^{13} W/cm^2$ (at the focus of the probe beam).

Figure 3 shows the TFISH amplification obtained for different iris diameters. As the iris diameter is reduced from 1.3mm to 0.3mm the TFISH signal increases by a factor of ~2 for all the measured probe pulse intensities and the amplification factor increases from 24±2 to 43±2. This ~2 fold increase in TFISH signal may result from THz diffraction and corresponding locally enhanced THz field and TFISH signal. This possibility was dismissed by replacing the aperture with a thin metal mesh (~250µm rectangular apertures) with no apparent difference. Thus, the ~2 fold increase obtained with reduced iris diameter is attributed to the improved THz beam blocking by the 0.3mm iris compared to the 1.3mm iris. Note that the TFISH enhancement does not depend on the probe intensity in the measured range. We also note that for the larger iris diameters, the amplification is fairly fixed with ~26-fold for 1mm aperture and ~24 fold for the 1.3mm aperture. Thus in the remaining of this paper we set the iris diameter to 1mm to avoid possible scattering of the probe by the iris edges and to simplify the beam alignment procedures.

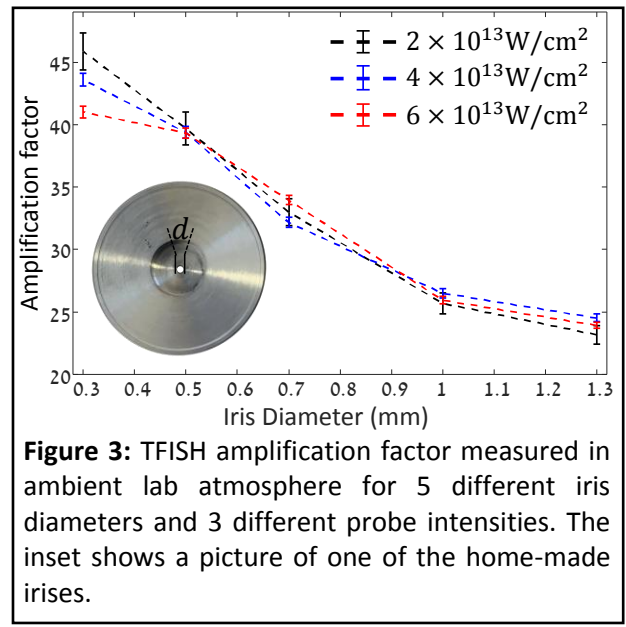

**Figure 3:** TFISH amplification factor measured in ambient lab atmosphere for 5 different iris diameters and 3 different probe intensities. The inset shows a picture of one of the home-made irises.

We conclude that by placing an aperture at the vicinity of the focal point we effectively reduce the interaction length by ~50% nevertheless obtain a few-ten-fold increase in the TFISH signal. Under perfect phase-matching conditions ($\Delta k \sim 0$, that can be satisfied at low density gas samples), the generated harmonic signal is expected to increase quadratically with the interaction length ($I_{TFISH} \propto L^2$). Thus, the observed iris-assisted TFISH enhancement may arise from the phase mismatch at the atmospheric air pressure of Figs.2,3. Different from the typically fixed non-linear coefficients of frequency doubling crystals (such as BBO, LiNbO$_3$ etc.) and their geometrical parameters, here, it is the THz field that governs the spatially (and temporally) dependent non-linear susceptibility throughout the interaction volume and the careful positioning of the iris that sets the length of the THz-induced non-linear gas medium.

**Phase matching condition**

In order to study the effect of phase mismatch on the TFISH signal in air, we performed a series of measurements with fixed probe intensity (20mW corresponding to $4 \times 10^{13} W/cm^2$ at the beam focus) and varying air pressures. Figure 4 depicts the peak TFISH signal intensity obtained with/without an iris for different gas pressures.

Under perfect phase matching conditions the TFISH intensity is expected to increase quadratically with the gas density ($I_{TFISH} \propto P^2$)[22]. Figure 4 shows the experimental TFISH results at varying air pressures, with and without the iris and at varying probe intensities ($2 - 1.2 \times 10^{13} W/cm^2$).

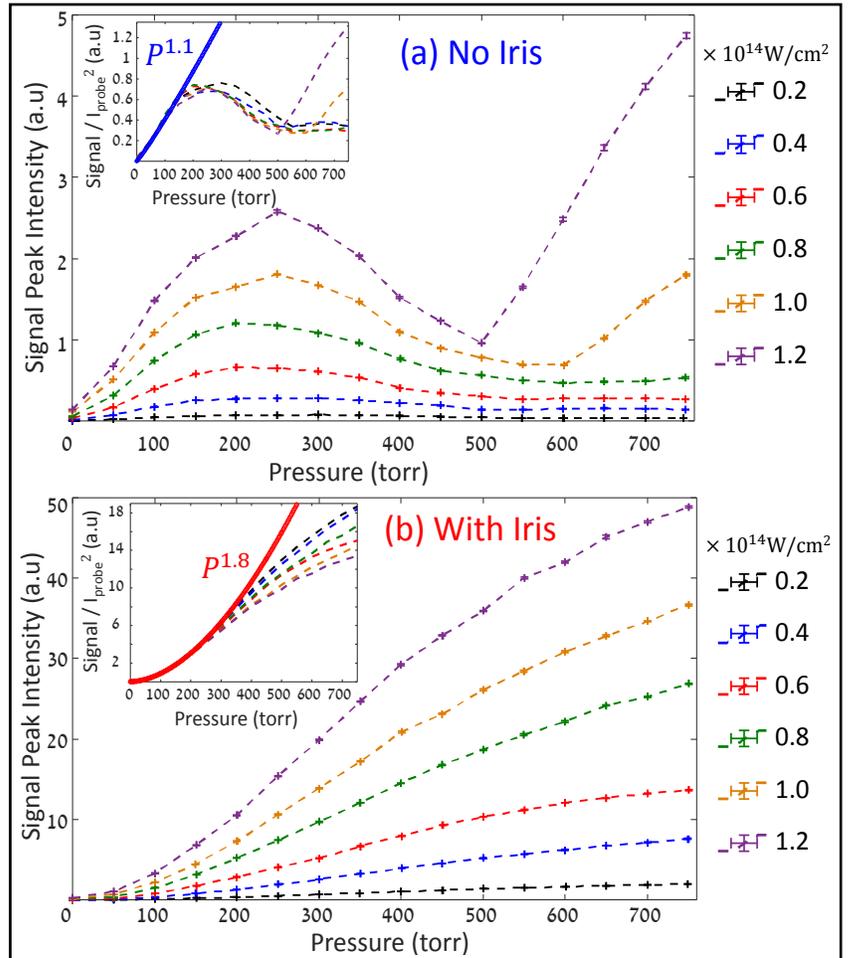

Fig.4a depicts the experimental TFISH signals obtained without the iris. At the low pressure region (P<200torr) the TFISH signal increases with pressure and the maximal TFISH signal is measured around 200-250torr for all of the probe intensities measured. The inset shows the same data, only normalized by the probe intensity squared. The agreement between the normalized data curves indicates the quadratic dependence of the TFISH signal on the probe intensity. However, the pressure dependence is found to severely deviate from quadratic as readily observed from the results. In fact, even at the low pressure region (P<~100torr) we get a dependence of $\propto P^{\sim 1.1}$ instead of the expected $\propto P^2$.

This dramatic deviation from quadratic dependence is due to the severe phase-mismatch between the probe (λ=800nm) and the generated signal (λ=400nm) which increases linearly with the gas density (pressure): $\Delta k = k_{2\omega} - 2k_\omega = -2\pi P \frac{n_0^{800nm} - n_0^{400nm}}{400nm}$ (2), as shown in [20]. However, the effect of phase mismatch also depends on the length of the interaction, $L$, given by: $I_{2\omega} \propto \left(\frac{\chi^{(2)} I_\omega}{\Delta k}\right)^2 \sin^2\left(\frac{\Delta k \cdot L}{2}\right)$ (3), as shown in [22,24]. For very small interaction length, where $\sin^2\left(\frac{\Delta k \cdot L}{2}\right) \approx \frac{1}{4} \Delta k^2 \cdot L^2$, eq.(3) becomes: $I_{2\omega} \propto \chi^{(2)^2} I_\omega^2 L^2$, similar to the case of perfect phase-matching conditions ($\Delta k = 0$).

**Figure 4:** Experimental TFISH signals obtained at varying air pressures and probe intensities (a) without the Iris and (b) with the Iris placed around the focus of the optical beam, inside the gas chamber. The insets in each figure depict the experimental data normalized by the probe intensity squared and fitted by a power law of $\propto P^{1.1}$ without the Iris and $\propto P^{1.8}$ with the Iris (see text).

Once the iris is positioned at the interaction volume (Fig.4b), the effect of phase-mismatch is significantly reduced owing to the smaller interaction length. Also here, the normalized data of Fig.4b (shown in the inset) indicates the expected quadratic dependence on the probe intensity, however, for the iris-restricted interaction length we obtain a power dependence of ~1.8, in good agreement with the expected quadratic dependence in the *P<300torr* range.

Figure 5 depicts the amplification factor (the ratio of the experimental data in 4b and 4a) as a function of gas pressure and probe intensity. At the low pressure regime (where $\Delta k$ is small) we obtain a small amplification factor of ~1.3 at 50torr that is extrapolated to ~1 (i.e. no amplification) for zero gas density. This result is easily explained by equation (2), where for low density (low P) the phase mismatch $\Delta k$ is also small and the non-linear TFISH process can be considered as perfectly phase-matched. Note that by positioning the iris in vicinity of the probe focus one would expect a decrease in the TFISH signal corresponding to an amplification factor < 1. In Fig.5 however, we find that the TFISH signal is fairly independent of the iris for low gas density (i.e. amplification factor ~1). This is attributed to our experimental procedure, where for each measurement we scan the position of the iris to obtain the maximal TFISH signal. This strategy partially compensates for nonlinear propagation effects such as self-focusing [27] that increases with both the gas density and the probe intensity.

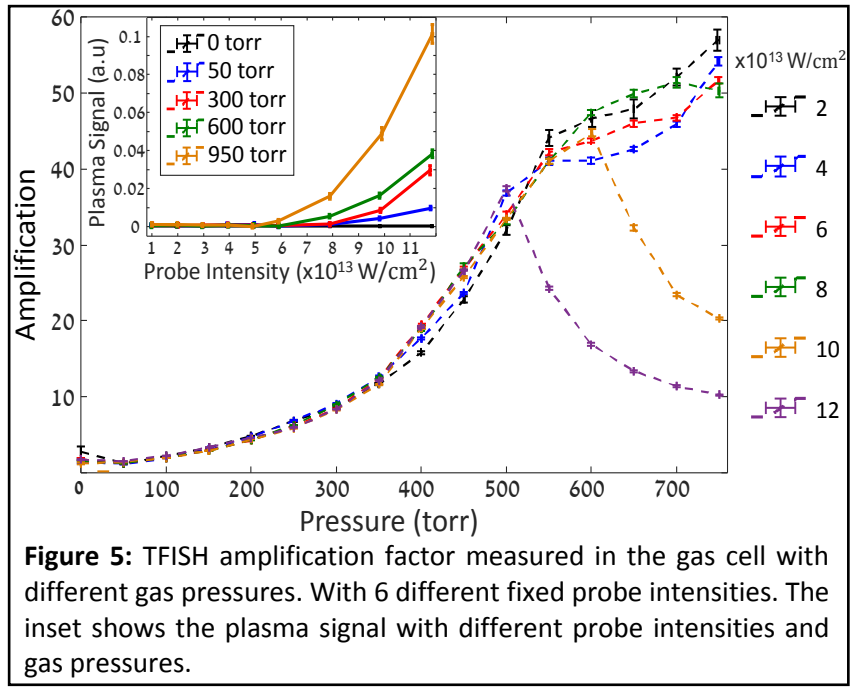

**Figure 5:** TFISH amplification factor measured in the gas cell with different gas pressures. With 6 different fixed probe intensities. The inset shows the plasma signal with different probe intensities and gas pressures.

As the gas density increases so does $\Delta k$ and gives rise to increased amplification factor when the iris is positioned at the interaction region as evident in Fig.5. In the region of ~50-500torr the increasing phase-mismatch is compensated by the truncation of the interaction length by the iris. This tendency persists at higher pressures (up to 760torr measured in Fig.5) for the low probe intensities but seem to break at higher probe intensities as evident from the sharp decrease in amplification factor for $1.0 \times 10^{14}\,\text{W/cm}^2$ (yellow line) and $1.2 \times 10^{14}\,\text{W/cm}^2$ (purple line) probe intensities above 600torr and 500torr respectively. The sharp decrease in amplification is due to the increase in TFISH signal obtained without the iris (see Fig.4a) and results from plasma generation via strong-field ionization of the gas. Note that with probe intensities up to $8 \times 10^{13}\,\text{W/cm}^2$ the tendency of the amplification factor persists throughout the entire range of gas densities measured, with an observed leveling at the higher densities, also attributed to initiation of plasma generation.

**The effect of laser-induced plasma**
The contribution of laser-induced plasma to the TFISH signal was previously described in ref.[17], where Dai et.al. categorize three regimes of TFISH detection: incoherent, hybrid and coherent TFISH signals governed by the probe intensity. With probe intensities $I_\omega > 5 \times 10^{14}\,W/cm^2$ the TFISH signal is heterodyned by a local oscillator (LO) field emanating from the white light generated via self-phase modulation [17,19], and was referred to as coherent TFISH detection. With our probe intensity $I_\omega < 1.2 \times 10^{14}\,W/cm^2$ the TFISH signal is well within the region of incoherent detection as observed by the fully homodyne time-resolved scans in Fig.2b. Our experimental results shown in Fig.4a (and correspondingly in Fig.5) unveil the contribution of the laser-induced plasma to the TFISH generation efficiency [28,29]. Different from [17] where the plasma provides a LO field, in our experiment, the LO field is negligibly small compared to the TFISH signal hence thwarting any detectable heterodyning of the signal. Nevertheless, for probe energies >30μJ (corresponding to $I_\omega > 5 \times 10^{13}\ W/cm^2$) we start detecting the signal coming from laser-induced plasma as shown in the inset of Fig.5 where we measured the plasma-induced white light at 400nm without the THz beam. Note that the signal scale in the inset is the same as in Fig.4a,b, namely, the pure laser-induced plasma signal is 1-2 orders of magnitude lower than the pure TFISH signals and therefore insufficient for heterodyning the TFISH signal. Nevertheless, the observed (weak) emission indicates the existence of plasma that increases with the probe intensity and with the gas pressure as expected. From all of the above we attribute the sharp increase in TFISH generation efficiency observed in Fig.4a to plasma-assisted phase-matching and possibly plasma enhanced third order susceptibility [30] that occurs within the incoherent regime of detection, yet dramatically contributing to increase the TFISH intensity. Note that the iris-assisted TFISH signal in Fig.4b is hardly affected by the generation of plasma. From the normalized data in the inset of Fig.4b we find that the Iris-assisted TFISH efficiency tends to decrease with the probe intensity, i.e. once the phase-mismatch is compensated by the iris – the contribution of the plasma (that slightly reduces the refractive index at the interaction region) is destructive in terms of the TFISH efficiency.

**Semi-infinite gas cells and effective interaction region**
On a broader perspective, the iris-assisted TFISH experiments described above are reminiscent of the semi-infinite gas cell (SIGC) geometry utilized in high-harmonic generation spectroscopy [31–34]. In SIGC, the gas occupies the volume between the focusing lens to a thin exit foil near the focus, after which the generated harmonics propagate in vacuum toward the detector. The enhanced efficiency of harmonic generation provided by the SIGC is attributed to the reduced re-absorption of the generated harmonic fields by the (very low) gas density after the focus. In our scheme, however, the second harmonic signal (400nm) does not benefit from propagation in vacuum since it is not re-absorbed by the gas (air). Instead, the THz field that enables the SHG is allowed to co-propagate with the optical probe until the iris that is positioned at the focus and provides the analogue of SIGC in the TFISH setup. As discussed above, the main reason for the iris-assisted TFISH amplification is due to the phase-mismatch imposed by the density of the gas at the interaction region. Thus in addition to providing a dramatically enhanced TFISH signal, the iris-assisted TFISH provides a controllable and elegant platform for studying nonlinear phase-matching phenomena; it enables clear distinction between re-absorption by the medium, geometric beam propagation effects and the ramifications of laser-induced plasma on the nonlinear optical process.

In what follows we apply the iris-assisted TFISH setup to characterize the interaction region of the TFISH process. We chose the simplest possible beam shape manipulation by truncation of the probe beam using a ring-actuated iris diaphragm.

The experiment is performed as follows: we set the probe beam diameter by varying the ring-actuated iris (Iris 1 in Fig.6) in the range 4-12mm. For each probe diameter we set the power transmitted through Iris 1 to 10mW (10μJ/pulse). Note that as the diameter of the iris-diaphragm is gradually reduced, the initial Gaussian intensity distribution evolves into a 'top-hat' shape. We fix the THz-probe delay at the peak intensity of the TFISH signal and for each truncated beam diameter we record the TFISH intensity as a function of Iris 2 position.

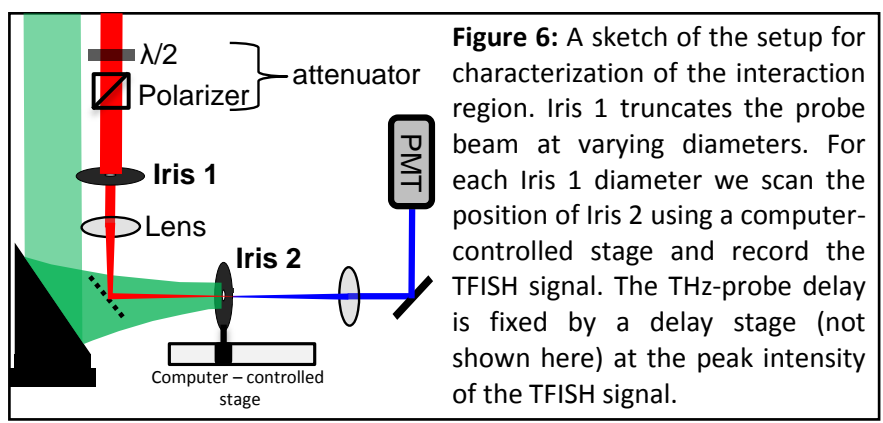

**Figure 6:** A sketch of the setup for characterization of the interaction region. Iris 1 truncates the probe beam at varying diameters. For each Iris 1 diameter we scan the position of Iris 2 using a computer-controlled stage and record the TFISH signal. The THz-probe delay is fixed by a delay stage (not shown here) at the peak intensity of the TFISH signal.

Figure 7 shows the experimental characterization of the effective interaction region. Iris 2 position 0 mm is set by the peak TFISH signal obtained with the full Gaussian probe beam (~12mm diameter). With the increasing truncation of the probe beam by Iris 1 (from 12mm to 4mm) the effective interaction region becomes wider as expected due to the increased Rayleigh range $z_R = \frac{\pi w_0^2}{\lambda}$, with $w_0 = \left(\frac{2\lambda}{\pi}\right)\left(\frac{f}{D}\right)$ where $D$ is the beam diameter at the plane of the lens with focal length $f$. Note the gradual shift of the interaction region toward the downstream direction (as observed by the red [6mm] and the blue [4mm] probe diameters). This results from the phase-mismatch between the 800nm and 400nm in the ambient atmosphere of our experiment (and does not occur in simulations with $\Delta k = 0$). The observed experimental trends are fully captured by our numerical simulations which are performed for pure Gaussian beams (clearly different from the actual intensity distribution of the truncated beams).

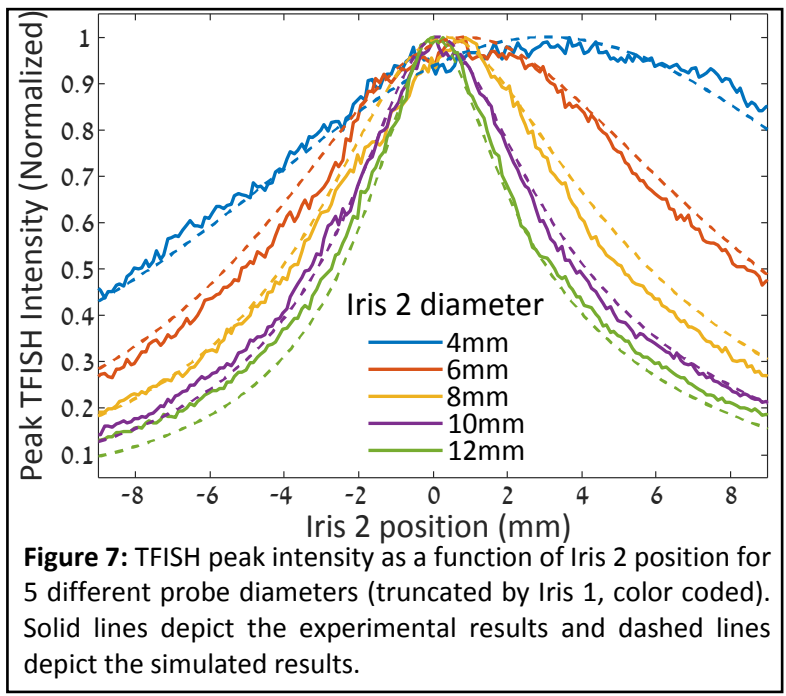

**Figure 7:** TFISH peak intensity as a function of Iris 2 position for 5 different probe diameters (truncated by Iris 1, color coded). Solid lines depict the experimental results and dashed lines depict the simulated results.

## Conclusions

In this work we have demonstrated the dramatic ramifications of phase-mismatch to the nonlinear TFISH generated in ambient air (that serves as the typical medium for TFISH characterization of THz fields). By restricting the interaction length of the two participating fields by simply placing an iris around the mutual focus of the fields, we obtained up to ~55 fold increase in the generated TFISH intensity. We have found that even at low probe intensities considered well within the region of incoherent TFISH detection, self-phase modulation and plasma generation manifest by enhanced TFISH efficiency. The demonstrated technique provides an experimental simulator for studying the effect of semi-infinite gas cells that are widely used in HHG with full control over the relevant experimental parameters such as the phase-matching conditions (by changing the gas type and/or density), beam geometries, interaction length (by varying the iris position), etc. The latter is utilized for experimental characterization of the effective interaction region for the TFISH process. Our observations are bound to affect and manifest in our current efforts to utilize the TFISH technique for probing the angular distribution of gas phase molecules and their coherent rotational dynamics. Lastly, the experimental system used in this experiment is the basic setup that is needed for TFISH detection in THz spectroscopy. The ability to increase the signal intensity by few-tens fold using a simple iris provides a practical solution for the detection of weaker THz fields and simplifies the alignment and calibration of the experimental apparatus.

**Financial support:** This work was supported by the Wolfson Foundation (Grants No. PR/ec/20419 and PR/eh/21797), the Israel Science Foundation – ISF (Grants No. 1065/14, 926/18, 2797/11 and by INREP—Israel National Research Center for Electrochemical Propulsion.

## References

1. H. Y. Hwang, S. Fleischer, N. C. Brandt, B. G. Perkins, M. Liu, K. Fan, A. Sternbach, X. Zhang, R. D. Averitt, and K. A. Nelson, "A review of non-linear terahertz spectroscopy with ultrashort tabletop-laser pulses," J. Mod. Opt. **62**(18), 1447–1479 (2015).
2. T. Kampfrath, K. Tanaka, and K. A. Nelson, "Resonant and nonresonant control over matter and light by intense terahertz transients," Nat. Photonics **7**(9), 680–690 (2013).
3. C. A. Schmuttenmaer, "Exploring Dynamics in the Far-Infrared with Terahertz Spectroscopy," Chem. Rev. **104**(4), 1759–1780 (2004).
4. D. Nicoletti and A. Cavalleri, "Nonlinear light–matter interaction at terahertz frequencies," Adv. Opt. Photonics **8**(3), 401 (2016).
5. D. Saeedkia, "Terahertz Photoconductive Antennas : Principles and Applications," 3326–3328 (2011).
6. X. Sun, F. Buccheri, J. Dai, and X.-C. Zhang, "Review of THz wave air photonics," in C. Zhang, X.-C. Zhang, H. Li, and S.-C. Shi, eds. (2012), **8562**, p. 856202.
7. M. C. Hoffmann, K.-L. Yeh, J. Hebling, and K. A. Nelson, "Efficient terahertz generation by optical rectification at 1035 nm," Opt. Express **15**(18), 11706 (2007).
8. R. Huber, A. Brodschelm, F. Tauser, and A. Leitenstorfer, "Generation and field-resolved detection of femtosecond electromagnetic pulses tunable up to 41 THz," Appl. Phys. Lett. **76**(22), 3191 (2000).
9. M. Shalaby and C. P. Hauri, "Demonstration of a low-frequency three-dimensional terahertz bullet with extreme brightness," Nat. Commun. **6**(1), 5976 (2015).
10. S. Keren-Zur, M. Tal, S. Fleischer, D. M. Mittleman, and T. Ellenbogen, "Generation of spatiotemporally tailored terahertz wavepackets by nonlinear metasurfaces," Nat. Commun. **10**(1), 1778 (2019).
11. L. Luo, I. Chatzakis, J. Wang, F. B. P. Niesler, M. Wegener, T. Koschny, and C. M. Soukoulis, "Broadband terahertz generation from metamaterials," Nat. Commun. **5**(1), 3055 (2014).
12. Q. Wu and X. C. Zhang, "Free-space electro-optic sampling of terahertz beams," Appl. Phys. Lett. **67**, 3523 (1995).
13. A. Nahata and T. F. Heinz, "Detection of freely propagating terahertz radiation by use of optical second-harmonic generation," Opt. Lett. **23**(1), 67 (1998).
14. D. J. Cook, J. X. Chen, E. A. Morlino, and R. M. Hochstrasser, "Terahertz-field-induced second-harmonic generation measurements of liquid dynamics," Chem. Phys. Lett. **309**, 221–228 (1999).
15. J. Chen, P. Han, and X.-C. Zhang, "Terahertz-field-induced second-harmonic generation in a beta barium borate crystal and its application in terahertz detection," Appl. Phys. Lett. **95**(1), 011118 (2009).
16. G. Kaur and X. C. Zhang, "Terahertz field induced second harmonic generation in biomolecules," in *IRMMW-THz 2011 - 36th International Conference on Infrared, Millimeter, and Terahertz Waves* (IEEE, 2011), pp. 1–1.
17. J. Dai, X. Xie, and X.-C. Zhang, "Detection of Broadband Terahertz Waves with a Laser-Induced Plasma in Gases," Phys. Rev. Lett. **97**(10), 103903 (2006).
18. X. Lu and X.-C. Zhang, "Terahertz Wave Gas Photonics: Sensing with Gases," J. Infrared, Millimeter, Terahertz Waves **32**(5), 562–569 (2011).
19. P. Babilotte, L. H. Coudert, F. Billard, E. Hertz, O. Faucher, and B. Lavorel, "Experimental and theoretical study of free induction decay of water molecules induced by terahertz laser pulses," Phys. Rev. A **95**(4), 043408 (2017).
20. I.-C. Ho, X. Guo, and X.-C. Zhang, "Design and performance of reflective ultra-broadband terahertz time-domain spectroscopy with air-biased-coherent-detection," in L. P. Sadwick and C. M. M. O'Sullivan, eds. (International Society for Optics and Photonics, 2010), p. 76010K.
21. H. He and X.-C. Zhang, "Analysis of Gouy phase shift for optimizing terahertz air-biased-coherent-detection," Appl. Phys. Lett. **100**(6), 061105 (2012).
22. X. Lu, N. Karpowicz, and X.-C. Zhang, "Broadband terahertz detection with selected gases," J. Opt. Soc. Am. B **26**(9), A66 (2009).
23. H. Lin, P. Braeuninger-Weimer, V. S. Kamboj, D. S. Jessop, R. Degl'Innocenti, H. E. Beere, D. A. Ritchie, J. A. Zeitler, and S. Hofmann, "Contactless graphene conductivity mapping on a wide range of substrates with terahertz time-domain reflection spectroscopy," Sci. Rep. **7**(1), 10625 (2017).
24. N. Karpowicz, X. Lu, and X.-C. Zhang, "Terahertz gas photonics," J. Mod. Opt. **56**(10), 1137–1150 (2009).
25. A. Dogariu, B. Goldberg, S. O'Byrne, and R. Miles, "Femtosecond Localized Electric Field Measurement in Gases via Second Harmonic Generation," in *Conference on Lasers and Electro-Optics* (OSA, 2017), p. FM4F.4.
26. S. Uusi-Heikkilä, D. Bierbach, J. Alós, P. Tscheligi, C. Wolter, and R. Arlinghaus, "Relatively large males lower reproductive success in female zebrafish," Environ. Biol. Fishes **101**(11), 1625–1638 (2018).
27. Boyd. R. W, *Nonlinear Optics* (2007).
28. K. Hartinger and R. A. Bartels, "Enhancement of third harmonic generation by a laser-induced plasma," Appl. Phys.

Lett. **93**(15), 151102 (2008).
29. C. Rodríguez, Z. Sun, Z. Wang, and W. Rudolph, "Characterization of laser-induced air plasmas by third harmonic generation," Opt. Express **19**(17), 16115 (2011).
30. S. Suntsov, D. Abdollahpour, D. G. Papazoglou, and S. Tzortzakis, "Filamentation-induced third-harmonic generation in air via plasma-enhanced third-order susceptibility," Phys. Rev. A **81**(3), 033817 (2010).
31. J. Peatross, J. R. Miller, K. R. Smith, S. E. Rhynard, B. W. Pratt, J. U. S T I N Peatross, J. U. L I A R Miller, K. R. S M I T H, and B. W. P R A T T, "Phase matching of high-order harmonic generation in helium-and neon-filled gas cells," J. Mod. Opt. **51**(16), 2675–2683 (2004).
32. M. Turner, N. Brimhall, M. Ware, and J. Peatross, "Simulated laser-pulse evolution for high-order harmonic generation in a semi-infinite gas cell," Opt. Express **16**(3), 1571 (2008).
33. L. Van Dao, S. Teichmann, and P. Hannaford, "Phase-matching for generation of few high order harmonics in a semi-infinite gas cell," Phys. Lett. A **372**(31), 5254–5257 (2008).
34. D. S. Steingrube, T. Vockerodt, E. Schulz, U. Morgner, and M. Kovačev, "Phase matching of high-order harmonics in a semi-infinite gas cell," Phys. Rev. A **80**(4), 043819 (2009).